\UseRawInputEncoding

\documentclass[letterpaper]{cas-dc}

\usepackage[numbers]{natbib}

\usepackage{amssymb,mathrsfs,bm} 
\usepackage{graphicx}

\usepackage[labelformat=simple]{subcaption}

\usepackage{enumitem}
\usepackage{float}

\usepackage{dcolumn}
\usepackage{siunitx}
\usepackage{pifont}

\usepackage{multirow}

\newcommand{\rd}{\mathrm{d}}

\usepackage{hyperref}
\hypersetup{
    colorlinks=true,
    linkcolor=blue,
    filecolor=magenta,
    urlcolor=cyan,
    citecolor=blue,
    breaklinks=true
}

\usepackage{tikz,etoolbox}
\newrobustcmd*{\mycirc}{\tikz{\filldraw[draw=cyan,fill=cyan!100] (0,0) circle [radius=0.075cm];}}
\newrobustcmd*{\mytriang}{\tikz{\filldraw[draw=magenta,fill=magenta!100] (-0.1,0) --(0.1cm,0.0) -- (0,0.15cm) -- (-0.1,0);}}

\allowdisplaybreaks

\begin{document}
\let\WriteBookmarks\relax
\def\floatpagepagefraction{1}
\def\textpagefraction{.001}

\shorttitle{Flow rate--pressure drop relations for slender compliant tubes} 

\shortauthors{X Wang et~al.}

\title[mode = title]{Flow rate--pressure drop relations for new configurations of slender compliant tubes arising in microfluidics experiments}

\tnotemark[1]
\tnotetext[1]{This work originates from research partially supported by the US National Science Foundation under grant No.\ CBET-1705637.}

\author[1]{Xiaojia~Wang}[orcid=0000-0002-3487-1600]
\ead{wang4142@purdue.edu}
\credit{Conceptualization, Methodology, Software, Validation, Formal analysis, Investigation, Data Curation, Writing - Original Draft, Visualization, Supervision}

\author[1]{Shrihari~D.\ Pande}[orcid=0000-0001-6962-8400]
\credit{Software, Validation, Investigation, Data Curation, Writing - Original Draft, Visualization}

\author[1]{Ivan~C.\ Christov}[orcid=0000-0001-8531-0531]
\cormark[1]
\ead{christov@purdue.edu}
\ead[url]{tmnt-lab.org}
\credit{Conceptualization, Methodology, Investigation, Writing - Original Draft, Supervision, Project administration, Funding acquisition}

\address[1]{School of Mechanical Engineering, Purdue University, West Lafayette, Indiana 47907, USA}

\cortext[cor1]{Corresponding author}

\begin{abstract}
We investigate the steady-state fluid--structure interaction between a Newtonian fluid flow and a deformable microtube in two novel geometric configurations arising in recent microfluidics experiments. The first configuration is a cylindrical fluidic channel surrounded by an annulus of soft material with a rigid outer wall, while the second one is a cylindrical fluidic channel extruded from a soft rectangular slab of material. In each configuration, we derive a mathematical theory for the nonlinear flow rate--pressure drop relation by coupling lubrication theory for the flow with linear elasticity for the inner tube wall's deformation. Using the flow conduit's axial slenderness and its axisymmetry, we obtain an analytical expression for the radial displacement in each configuration from a plane-strain configuration. The predicted displacement field, and the resulting closed-form flow rate--pressure drop relation, are each validated against three-dimensional direct numerical simulations via SimVascular's two-way-coupled fluid--structure interaction solver, svFSI, showing good agreement. We also show that weak flow inertia can be easily incorporated in the derivation, further improving the agreement between theory and simulations for larger imposed flow rates.
\end{abstract}




\maketitle

\section{Introduction}
\label{sec:intro}

\emph{Microfluidics}, which concerns the transport of small volumes of fluids at microscopic scales, has emerged as a fundamental research field over the last several decades \cite{NW06,SQ05}. Fluid flows at small scales involve the coupling of physical effects that are often not observable at larger scales \cite{B08}, and necessitate updating (or redevelopment) of aspects of the basic continuum theories \cite{St15,St17}. More specifically, recently, there has been a growing interest in the topic of \emph{soft hydraulics} \cite{C21}, \textit{i.e.}, small-scale flows in compliant conduits, due to the wealth of mechanical \cite{DS15}, biological \cite{F97}, physical \cite{B08} and technological \cite{SGOG20,PBBK22} problems involving such fluid--structure interactions. Typically, the pressure drop required to maintain a steady flow within compliant conduits varies nonlinearly with the flow rate, deviating from the classic Poiseuille (or Hagen--Poiseuille) law for rigid conduits \cite{GEGJ06,C21}. Using perturbation methods, previous studies have successfully derived three-dimensional solutions, leading to predictive theories that quantify this nonlinear flow rate--pressure drop relation in rectangular microchannels \cite{CCSS17,SC18,WC19,BSC22} and in axisymmetric microtubes with thin walls \cite{EG14,BBG17,AC18b,PV22}. 

Recently, two new configurations of compliant, axisymmetric cylindrical geometries have been considered in experiments: (\textit{a}) a thick, elastic annulus constrained between a fluidic channel and a rigid outer cylinder \cite{KMZ21}, and (\textit{b}) a tall and wide rectangular block of elastic material from which a cylindrical tube has been extruded creating a fluidic channel \cite{RCDC18,KDMCC20}, as shown in Fig.~\ref{fig:geometries}. However, a complete theory of the viscous fluid--structure interaction in these novel compliant ``microtube'' configurations is lacking. To this end, in this work, we derive the pressure--deformation and flow rate--pressure drop relations for Newtonian fluid flow through these configurations. We show that our theory, derived using the lubrication approximation and linear elasticity for slender structures, agrees with three-dimensional, two-way coupled direct numerical simulations performed using the open-source software package SimVascular \cite{Upde17,Lan18}.

Specifically, in section~\ref{subsec:fluid}, we review lubrication theory for axisymmetric flow in a slender cylindrical flow conduit. In section~\ref{subsec:solid}, the pressure--deformation relation for each configuration is derived from the theory of linear elasticity. Then, in section~\ref{subsec:couple}, the closed-form flow rate--pressure drop relation is obtained. In section~\ref{sec:num}, we describe the methodology for performing direct numerical simulations to validate our theory. Finally, in section~\ref{sec:results}, we discuss the comparisons between theory and simulation.

\section{Problem formulation and mathematical analysis}

\begin{figure*}[ht!]
    \centering
    \includegraphics[width=0.9\textwidth]{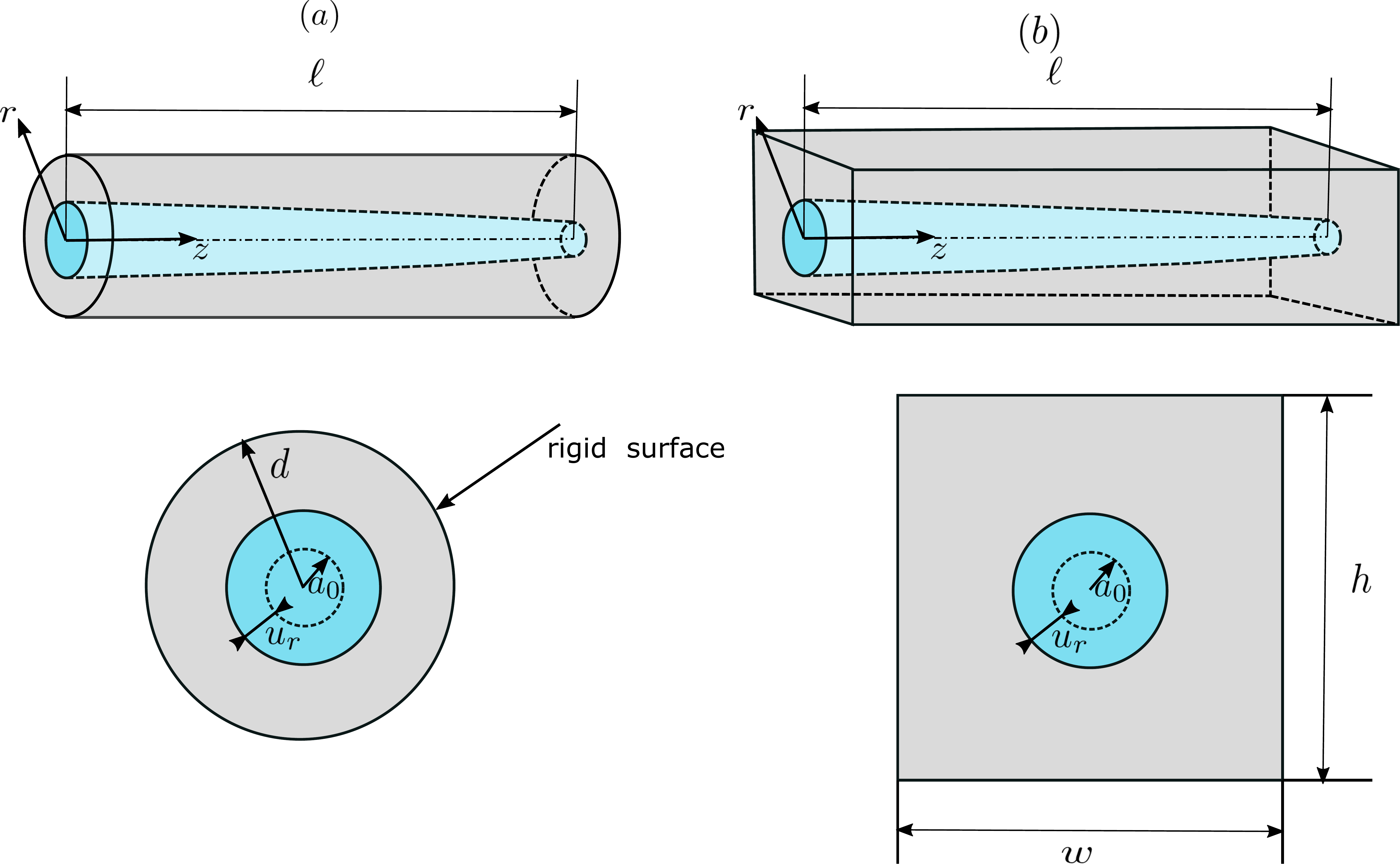}
    \caption{Schematics of two configurations of a Newtonian fluid flowing through compliant fluidic conduits, with the first row showing each three-dimensional configuration and the second row showing cross-sectional views. (\textit{a}) The fluidic conduit is surrounded by an annulus of soft material with a rigid outer wall (no-displacement condition at $r=d$). (\textit{b}) The fluidic conduit is a cylindrical tube extruded from a soft material contained in a large rectangular slab (stress-free conditions at $x=\pm w/2$ and $y=\pm h/2$), with $w \gg a_0$ and $h \gg a_0$.}
    \label{fig:geometries}
\end{figure*}

We study the steady fluid--structure interaction between a viscous flow and an elastic confining structure in two geometric configurations of recent experimental interest depicted in Fig.~\ref{fig:geometries}. In the undeformed state, each fluidic conduit (tube) has a radius $a_0$ and axial length $\ell$. For both configurations, we assume the conduit is slender, such that $a_0\ll \ell$; \textit{i.e.}, the aspect ratio $\epsilon=a_0/\ell \ll 1$. We assume that the cross-section of the flow conduit remains circular upon deformation. Specifically, the deformation $\bm{u}$ remains axisymmetric, such that $\bm{u}=u_r(r,z)\bm{e}_r$, even after it expands due to the flow within. This assumption is kinematic in nature, based on the expectation of a state of plane strain (with small and/or negligible axial displacement along $\bm{e}_z$) in a long slender structure (see, \textit{e.g.}, \cite{WC19}). Thus, in cylindrical coordinates, the deformed fluid domain is $\{(r,\theta,z)\ |\ 0\leq r\leq a_0 + u_r,\ 0\leq\theta<2\pi,\ 0\leq z\leq \ell \}$. In what follows, we denote by $a(z) = a_0 + u_r(a_0,z)$ the deformed radius of the tube.

\subsection{Fluid mechanics}
\label{subsec:fluid}

Similar to the deformation, we  assume that the flow field $\bm{v}$ and the pressure $p$ satisfy the assumption of axisymmetry without swirl, such that  $\bm{v} = v_r(r,z)\bm{e}_r+v_z(r,z)\bm{e}_z$, $\partial\bm{v}/\partial \theta =0$, and $\partial p/\partial \theta=0$. Then, for a Newtonian fluid of density of $\rho_f$ and  dynamic viscosity of $\mu_f$, the incompressible Navier--Stokes equations at steady state take the form \cite{W06_book}:
\begin{subequations}\label{iNS}
\begin{equation}
    \underbrace{\frac{1}{r}\frac{\partial}{\partial r}(r v_r)}_{\mathcal{O}(1)} + \underbrace{\frac{\partial v_z}{\partial z}}_{\mathcal{O}(1)} =  0, \label{com}
\end{equation}
\begin{multline}
    \underbrace{\rho_f v_r\frac{\partial v_r}{\partial r}}_{\mathcal{O}(\epsilon^3 Re)} + \underbrace{\rho_f v_z\frac{\partial v_r}{\partial z}}_{\mathcal{O}(\epsilon^3 Re)} \\ 
    = \underbrace{\mu_f \frac{\partial}{\partial r}\left[ \frac{1}{r}\frac{\partial}{\partial r}(r v_r) \right]}_{\mathcal{O}(\epsilon^2)} + \underbrace{\mu_f\frac{\partial^2 v_r}{\partial z^2}}_{\mathcal{O}(\epsilon^4)} - \underbrace{\frac{\partial p}{\partial r}}_{\mathcal{O}(1)}, \label{colm-r}
\end{multline}
\begin{multline}
    \underbrace{\rho_f v_r\frac{\partial v_z}{\partial r}}_{\mathcal{O}(\epsilon Re)} + \underbrace{\rho_f v_z\frac{\partial v_z}{\partial z}}_{\mathcal{O}(\epsilon Re)} \\ 
    = \underbrace{\mu_f \frac{1}{r}\frac{\partial}{\partial r}\left( r\frac{\partial v_z}{\partial r}\right)}_{\mathcal{O}(1)} + \underbrace{\mu_f\frac{\partial^2 v_z}{\partial z^2}}_{\mathcal{O}(\epsilon^2)} - \underbrace{\frac{\partial p}{\partial z}}_{\mathcal{O}(1)}, \label{colm-z}
\end{multline}\label{eq:iNS}\end{subequations}
where $Re=\rho_f\mathcal{V}_c a_0/\mu_f$ is the  Reynolds number. In Eqs.~\eqref{eq:iNS}, the order-of-magnitude of each term is listed underneath, based on the scales from Table~\ref{tab:scales}.

\begin{table}[width=.9\linewidth,cols=4,pos=h]
    \caption{Scales for the variables in the incompressible Navier--Stokes equations \eqref{iNS}.}
    \begin{tabular*}{\tblwidth}{@{} L|CCCCCC@{} }
    \toprule    
      Variable  & $r$ & $z$ & $v_r$ & $v_z$ & $p$ \\
    \midrule
      Scale  & $a_0$ & $\ell$ & $\epsilon\mathcal{V}_c$ & $\mathcal{V}_c$ & $\mathcal{P}_c = \frac{\mu_f \mathcal{V}_c \ell }{a_0^2}$\\
    \bottomrule
    \end{tabular*}
    \label{tab:scales}
\end{table}

In Table~\ref{tab:scales}, we have chosen $\epsilon\mathcal{V}_c$ and $\mathcal{V}_c$ as the characteristic scales for radial velocity $v_r$ and axial velocity $v_z$, respectively, in order to ensure a balance in the conservation of mass of Eq.~\eqref{com}. We have then chosen the viscous characteristic pressure $\mathcal{P}_c$ in terms of $\mathcal{V}_c$ to ensure a balance between the viscous forces and the pressure gradient in Eq.~\eqref{colm-z}. We are interested in the flow-controlled regime, in which the volumetric flow rate, $q$, is prescribed at the conduit's inlet, leading to the choice $\mathcal{V}_c = q/(\pi a_0^2) $, so that $\mathcal{P}_c = \mu_f q \ell /(\pi a_0^4)$ and $Re = \rho_f q /(\pi a_0 \mu_f)$.


Next we assume $\epsilon Re\ll 1$, which is the well known \emph{lubrication approximation} \cite{W06_book,B08}. The inertia of the flow is then negligible, and we are interested in the leading-order-in-$\epsilon$ solution to the hydrodynamic equations. Eq.~\eqref{colm-r} becomes $\partial p/\partial r = 0$ at leading order, which implies that $p=p(z)$ only. Then, the leading-order solution of Eq.~\eqref{colm-z} is
\begin{equation}\label{vz-O1}
    v_z(r,z) = \frac{1}{4\mu_f}\frac{\rd p}{\rd z}\left(r^2-a^2\right),
\end{equation}
where we have imposed the no-slip boundary condition, $v_r(a,z)=0$, at the deformed fluid--solid interface, $r=a(z)$.

Integrating Eq.~\eqref{vz-O1} over the deformed cross-section, we can relate the flow rate $q$ to the pressure gradient $\rd p/\rd z$ as
\begin{equation}\label{q-dp}
    q = \int_0^{2\pi} \!\! \int_0^{a(z)}  v_z(r,z) r\, \rd r \, \rd \theta = -\frac{\pi a^4(z)}{8\mu_f}\frac{\rd p}{\rd z},
\end{equation}
which is, of course, Poiseuille's law for a deformed tube. 

Finally, observe that we can relate $v_z$ to $q$ via Eq.~\eqref{vz-O1} and Eq.~\eqref{q-dp} by eliminating $\rd p/\rd z$. This relation can be used to obtain a leading-order-in-$\epsilon$ theory valid up to $\epsilon Re=\mathcal{O}(1)$. The details of this derivation are given in Appendix~\ref{app:a} for the interested reader.

\subsection{Solid mechanics}
\label{subsec:solid}

Eq.~\eqref{q-dp} is a first-order ordinary differential equation (ODE) for $p(z)$. In order to solve it, and obtain the relation between $q$ and $p$, we need to first obtain an expression for the deformed radius $a(z)$. To this end, we assume the deformation gradient in the solid wall to be small, so that the theory of linear elasticity applies \cite{Saad20}. Then, since the cross-sectional dimension of the tube is much smaller than its length (\textit{i.e.}, $\epsilon\ll1$), the dominant balance of stresses occurs in the conduit's cross-section, reducing the three-dimensional (3D) elasticity problem to a two-dimensional (2D) plane strain problem in the $(r,\theta)$ plane. As mentioned above, we assume the deformation is in the radial direction and the displacement field is axisymmetric. Thus, the deformed radius can be written as $a(z)=a_0+u_r(a_0, z)$. 

Under the above assumptions, the relation between the dominant stress components and the strains can be written in terms of just $u_r$ \cite{Saad20}:%
\begin{subequations}\label{cauchy-stresses}
\begin{align}
    \sigma_{rr} &= \lambda \frac{1}{r}\frac{\partial}{\partial r}(r u_r) + 2G\frac{\partial u_r}{\partial r},\\ \sigma_{\theta\theta} &= \lambda \frac{1}{r}\frac{\partial}{\partial r}\left( r u_r \right) + 2G\frac{u_r}{r},
\end{align}
\end{subequations}
where $\lambda$ and $G$ are the two Lam\'e constants in the linear stress-strain relation of an isotropic elastic solid. For such plane strain configurations, as shown in the bottom row of Fig.~\ref{fig:geometries}, Cauchy's balance of linear momentum (neglecting any body forces) becomes \cite{Saad20}:
\begin{equation}\label{cauchy-eq}
    \frac{1}{r}\frac{\rd}{\rd r}\left( r\sigma_{rr} \right)-\frac{\sigma_{\theta\theta}}{r} = 0.
\end{equation}
Substituting Eqs.~\eqref{cauchy-stresses} into Eq.~\eqref{cauchy-eq} and rearranging, we obtain
\begin{equation}\label{ur-eq}
    \frac{1}{r}\frac{\partial^2}{\partial r^2}\left( r u_r \right) - \frac{1}{r^2}\frac{\partial}{\partial r}\left( r u_r \right) = 0.
\end{equation}

Eq.~\eqref{ur-eq} can be solved by separation of variables. Let $u_r(r,z) = f(r)g(z)$ and substitute into Eq.~\eqref{ur-eq}. It is easy to show that the general solution for $f(r)$ is 
\begin{equation}\label{f-sol}
    f(r) = c_1 r + \frac{c_2}{r},
\end{equation}
where $c_1$ and $c_2$ are two constants of integration to be determined from suitable boundary conditions.

In configuration I, the outer surface rigid at $r=d$ restricts the displacement, $u_r(d, z)=0$, while the radial normal stress at the fluid--solid interface matches the hydrodynamic pressure \cite{EG14,BBG17,AC18b,PV22},  $\sigma_{rr}(a_0,z)=-p(z)$. Imposing these boundary conditions yields $c_1=-c_2/d^2$ and $g(z)=p(z)$, with
\begin{equation}
    c_2 = \frac{a_0^2}{2G\left[1+\left(\frac{\lambda+G}{G}\right)\frac{a_0^2}{d^2}\right]} = \frac{a_0^2}{2G\left[1+\left(\frac{1}{1-2\nu}\right)\frac{a_0^2}{d^2}\right]},
\end{equation}
where we have used the identity $(\lambda+G)/G = 1/(1-2\nu)$ for an isotropic linearly elastic solid, with $\nu$ being the Poisson's ratio. Finally, the radial displacement $u_r$ is found to be
\begin{equation}\label{ur-sol-a}
    u_r(r,z) = \frac{a_0^2}{2G\left[1+\frac{(a_0/d)^2}{1-2\nu}\right]}\left(\frac{1}{r}-\frac{r}{d^2}\right)p(z).
\end{equation}
Accordingly, for configuration~I,
\begin{equation}\label{a-sol-a}
\begin{aligned}
    a(z) &= a_0 + u_r(a_0,z) \\
    &= a_0 \Bigg\{ 1 + \underbrace{\frac{1}{2G\left[1+\frac{(a_0/d)^2}{1-2\nu}\right]}\left(1-\frac{a_0^2}{d^2}\right)}_{1/k} p(z)\Bigg\}.
\end{aligned}
\end{equation}
Here, for convenience of notation, we have introduced the constant $k$. Its physical relevance is discussed below.

In configuration II, we still have $\sigma_{rr}(a_0,z)=-p(z)$ on the fluid--solid interface, but now we do not have a confinement that imposes a no-displacement condition at $r=d$. Instead, to find the unique solution of the elasticity problem, we require that the stress field decay away from the fluid--solid interface: $\lim_{r\to\infty}\sigma_{rr}(r,z)=0$. This boundary condition then gives $c_1=0$, thus
\begin{equation}\label{ur-sol-b}
    u_r(r,z) = \frac{a_0^2}{2G r}p(z).
\end{equation}
Note that this solution was also obtained in \cite[Eq.~(8)]{RCDC18}, however the prefactor of $1/4$ therein should be corrected to $1/2$. 
The deformed radius is now 
\begin{equation}\label{a-sol-b}
\begin{aligned}
    a(z) &= a_0 + u_r(a_0, z)\\
    &= a_0\Bigg[ 1 + \underbrace{\frac{1}{2G}}_{1/k} p(z) \Bigg].
\end{aligned}
\end{equation}
Note that Eq.~\eqref{a-sol-b} can be obtained from Eq.~\eqref{a-sol-a} by also taking $d\to\infty$ (\textit{i.e.}, the rigid outer surface is located far away compared to the initial radius, $d\gg a_0$).

In both configurations, the fluid--solid interface's displacement $u_r(z)$ in Eq.~\eqref{ur-sol-a} and Eq.~\eqref{ur-sol-b} is linearly proportional to the local pressure $p(z)$, at a given streamwise location $z$, with the proportionality constant denoted as $1/k$ in Eq.~\eqref{a-sol-a} and Eq.~\eqref{a-sol-b}. Thus, $k$ can be interpreted as the \emph{effective stiffness} of the fluid--solid interface, in the sense of a Winkler foundation \cite{DMKBF18}, \emph{but without having assumed such a model} for the solid mechanics. Importantly, this derivation shows that the different geometrical and material properties of the two tube configurations considered (as well as the different stress boundary conditions) give rise to different stiffness expressions.

\subsection{Fluid--solid coupling}
\label{subsec:couple}

Substituting $a(z)$ from either Eq.~\eqref{a-sol-a} or \eqref{a-sol-b} into Eq.~\eqref{q-dp}, we can obtain a relation between $q$ and $p$ for both geometric configurations considered. At steady state $q$ is constant throughout the tube, and solving the ODE for $p(z)$ is straightforward. Imposing gauge pressure at the outlet, \textit{i.e.}, $p(\ell) = 0$, as is common in experiments, we obtain
\begin{equation}\label{q-p-1}
    (\ell -z )q=\frac{\pi a_0^4 k}{40\mu_f}\left\{[1+p(z)/k]^5-1\right\},
\end{equation}
which can be easily inverted to find the pressure as
\begin{equation}\label{q-p-2}
    p(z) = k\left\{ \left[ \frac{40(\ell-z)\mu_f q}{\pi a_0^4 k} +1 \right]^{1/5} -1\right\}.
\end{equation}
Note that Eq.~\eqref{q-p-1} and Eq.~\eqref{q-p-2} are valid for both geometric configurations, taking the expression for $k$ to be the one given in Eq.~\eqref{a-sol-a} or Eq.~\eqref{a-sol-b}, respectively. 

Obviously, a non-zero outlet pressure, $p(\ell)=p_{\mathrm{out}}$, can also be imposed. Modifying Eq.~\eqref{q-p-1} and Eq.~\eqref{q-p-2} for this case is a straightforward exercise left to the reader.

The pressure drop is defined as $\Delta p:=p(0)-p(\ell)$. Then, the flow rate--pressure drop relation is found by evaluating  Eq.~\eqref{q-p-2} at $z=0$ for the chosen geometric configuration. Finally, recall that both Eq.~\eqref{q-p-1} and Eq.~\eqref{q-p-2} are valid only for $\epsilon Re \to 0$. For completeness, in Appendix~\ref{app:a}, we also derive a flow rate--pressure drop relation for $\epsilon Re = \mathcal{O}(1)$ (see Eq.~\eqref{q-p-Re}).

\begin{table*}[width=.9\textwidth,cols=4,pos=h]
    \caption{Geometrical and material properties for the fluid--structure interaction problems solved for the two configurations shown in Fig.~\ref{fig:geometries}.}
    \centering
    \begin{tabular*}{\tblwidth}{@{} LLLLL@{} }
    \toprule
    Quantity & Variable & Configuration~I & Configuration~II & Units \\
    \hline
    Fluidic domain (tube's) length & $\ell$ & $2.5$ & $2.5$ & \si{\milli\meter} \\
    Fluidic domain (tube's) undeformed radius & $a_0$ & $25$ & $25$ & \si{\micro\meter} \\
    Outer/confinement radius & $d$  &  $50$, $225$ & --  & \si{\micro\meter} \\
    Outer/confinement width & $w$ & -- & $250$ & \si{\micro\meter}\\
    Outer/confinement height & $h$ & -- & $250$ & \si{\micro\meter}\\
    Young's modulus of wall material & $E$ & $0.5$ & $0.5$ & \si{\mega\pascal}\\
    Poisson's ratio of wall material & $\nu$ & $0.46$ & $0.46$ & -- \\
    Corresponding shear modulus, $E/[2(1+\nu)]$ & $G$ & $0.1712$ & $0.1712$ & \si{\mega\pascal}\\
    Fluid viscosity & $\mu_f$ & $0.89\times 10^{-3}$ & $0.89\times 10^{-3}$ & \si{\pascal\second}\\
    Fluid density & $\rho_f$ & $998$ & $998$ & $\si{\kilogram\per\meter\cubed}$\\
    \bottomrule
    \end{tabular*}
    \label{table:param}
\end{table*}

\section{Numerical simulation methodology}
\label{sec:num}

We performed 3D, two-way coupled direct numerical simulations using the open-source software package SimVascular \cite{Upde17,Lan18}. SimVascular has a fluid--structure interaction solver, `svFSI' \cite{svFSI}, which employs an arbitrary Lagrangian--Eulerian framework within the finite-element method to solve the coupled 3D equations of incompressible flow and elasticity. To test the small-deformation assumption made in our theory, we allowed for large deformation of the solid in the simulations by using the `Saint-Venant--Kirchhoff' constitutive model implemented in svFSI. 

First, using the commercial computer-aided engineering software ANSYS, we generated conforming, unstructured fluid and solid meshes of both microtube configurations depicted in Fig.~\ref{fig:geometries}. The fluid and the solid mesh were exported separately and independently converted to {\tt .vtu} files, which are suitable for svFSI simulations. Then, in svFSI, the fluid and the solid domain were assigned constant material properties as per Table~\ref{table:param}, comparable to experiments. 

{A steady solver is not implemented in svFSI, therefore we performed unsteady simulations starting with initial conditions corresponding to the fluid and solid being at rest, with the tube's initial state being of uniform radius $a_0$. At the fluid domain's inlet, a fully-developed parabolic (Poiseuille) velocity profile matched to flow rates of $q=50$, $100$, $125$, $175$, $300$, and $400~\si{\micro\litre\per\minute}$ was imposed at $t=0^+$. Meanwhile, at the fluid domain's outlet, the pressure was set to zero (gauge pressure), consistent with the theory. For both configurations, the solid domain at the inlet and the outlet was restricted by imposing a Dirichlet boundary condition of zero displacement. The outer wall in configuration~I was also restricted to have zero displacement, while the outer wall of configuration~II was set to be stress-free by imposing a Neumann boundary condition. We marched the simulations to steady state with a time step of $\Delta t = 10^{-5}~\si{\sec}$. The simulations were performed on a computational cluster. The transients `die out,' and a steady state was typically achieved, after $\approx 1\,000$ time steps. We determined that a steady state was achieved when $|\Delta p^{n+1}-\Delta p^n|/\Delta p^n < 0.01\%$, where $\Delta p^n$ represents the pressure drop at the $n$th time step. The pressure drop was computed by taking the area-averaged pressure at the fluid domain's inlet.}

To verify our simulations, we determined an appropriate grid size to use via a grid-independence study at a flow rate of $400~\si{\micro\litre\per\minute}$. Three different meshes, with the total number of elements roughly doubling between each ($642\,201$, $1\,329\,806$, $3\,388\,569$, respectively) were employed. The pressure drop, computed as described above, was compared across these three meshes. The relative error between the coarse and medium mesh was found to be $1.01\%$, while the relative error between the medium and fine mesh was found to be $3.05\%$, demonstrating grid convergence of the numerical solution. Since the relative error was small in all cases, the coarsest mesh was used for all simulations reported in this work to minimize computational time.

\section{Results and discussion}
\label{sec:results}

\begin{figure*}[ht]
    \centering
    \begin{subfigure}[b]{0.49\textwidth}
        \caption{}
        \includegraphics[width=\textwidth]{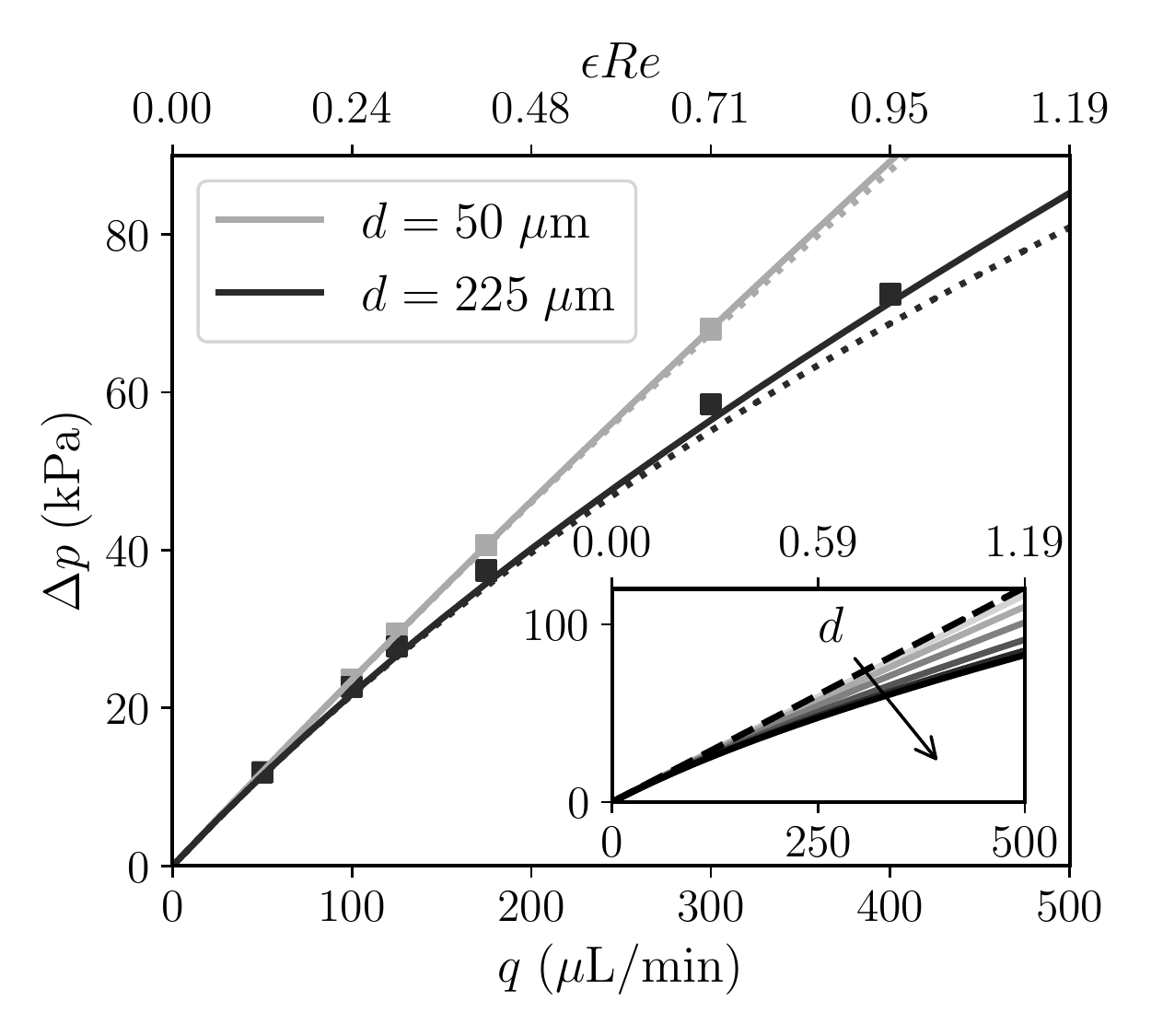}
        \label{subfig:config_a-q-dp}
    \end{subfigure}
    \begin{subfigure}[b]{0.49\textwidth}
    \centering
        \caption{}
        \includegraphics[width=.7\textwidth]{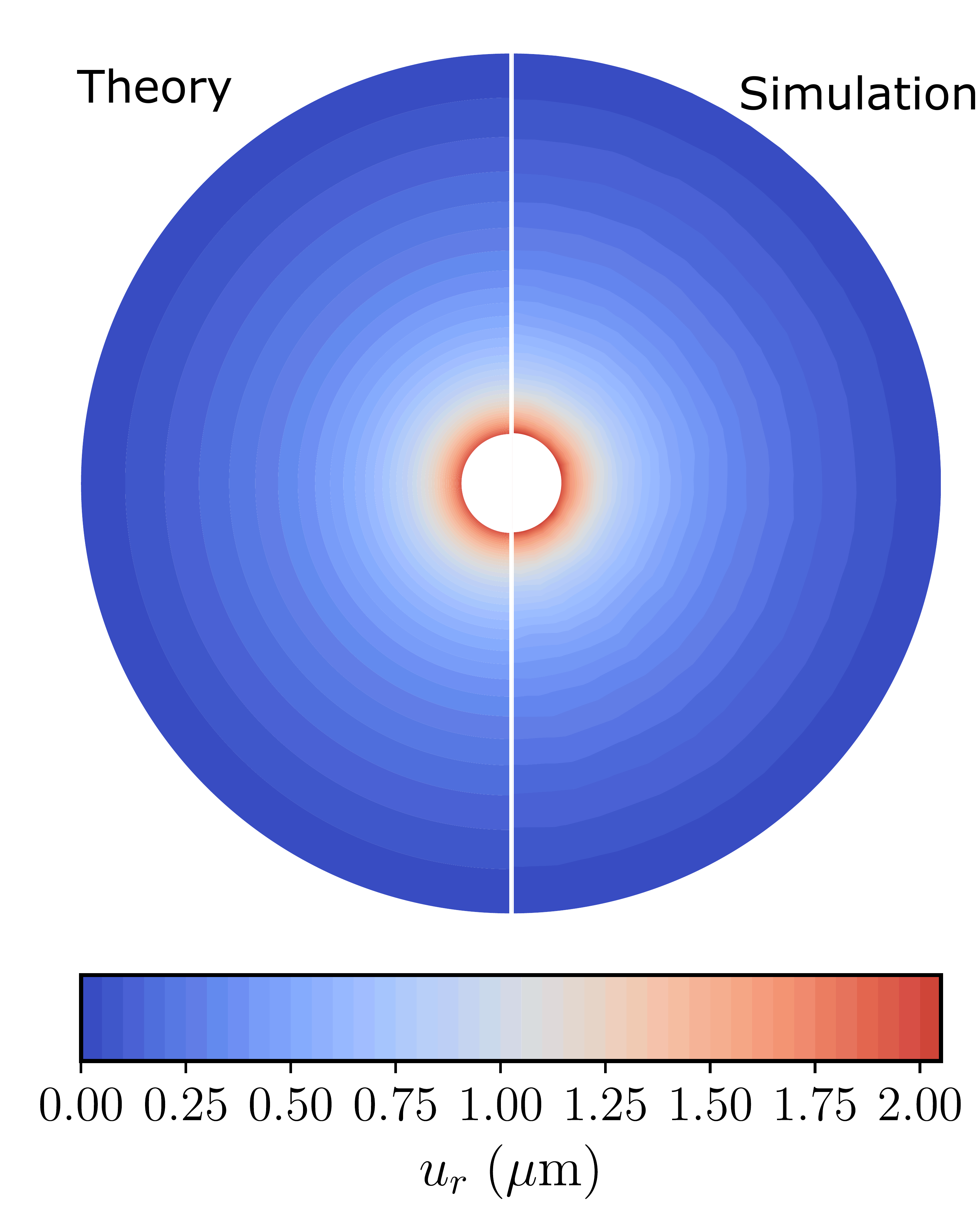}
        \label{subfig:config_a-cross}
    \end{subfigure}
    \caption{Comparison between the theory and the 3D direct simulation for configuration~I. (\textit{a}) Predicted flow rate--pressure drop relation. The solid and dotted curves represent $\Delta p$ from Eq.~\eqref{q-p-Re} and  Eq.~\eqref{q-p-2}, respectively. The symbols represent the simulation data. The inset in (\textit{a}) shows the $q$--$\Delta p$ relation predicted by Eq.~\eqref{q-p-Re} for $d=35,\ 50,\ 75,\ 125,\ 225,\ 425~\si{\micro\meter}$, {while the dashed line represents the linear Hagen--Poiseuille law ($q$--$\Delta p$ relation) for a rigid tube (\textit{i.e.}, $\Delta p=8\mu_f q\ell/(\pi a_0^4)$).} (\textit{b}) The radial displacement $u_r$ of the tube cross-section at $z=0.1\ell$, for $d=225~\si{\micro\meter}$ and $q=175~\si{\micro\litre\per\minute}$. The left half of the figure is plotted using the theory from Eq.~\eqref{q-p-Re} and Eq.~\eqref{ur-sol-a}, while the right half of the figure is plotted from the results of a SimVascular simulation.} 
    \label{fig:config_a}
\end{figure*}

\begin{figure*}[ht]
    \centering
    \begin{subfigure}[b]{0.49\textwidth}
    \centering
        \caption{}
        \includegraphics[width=\textwidth]{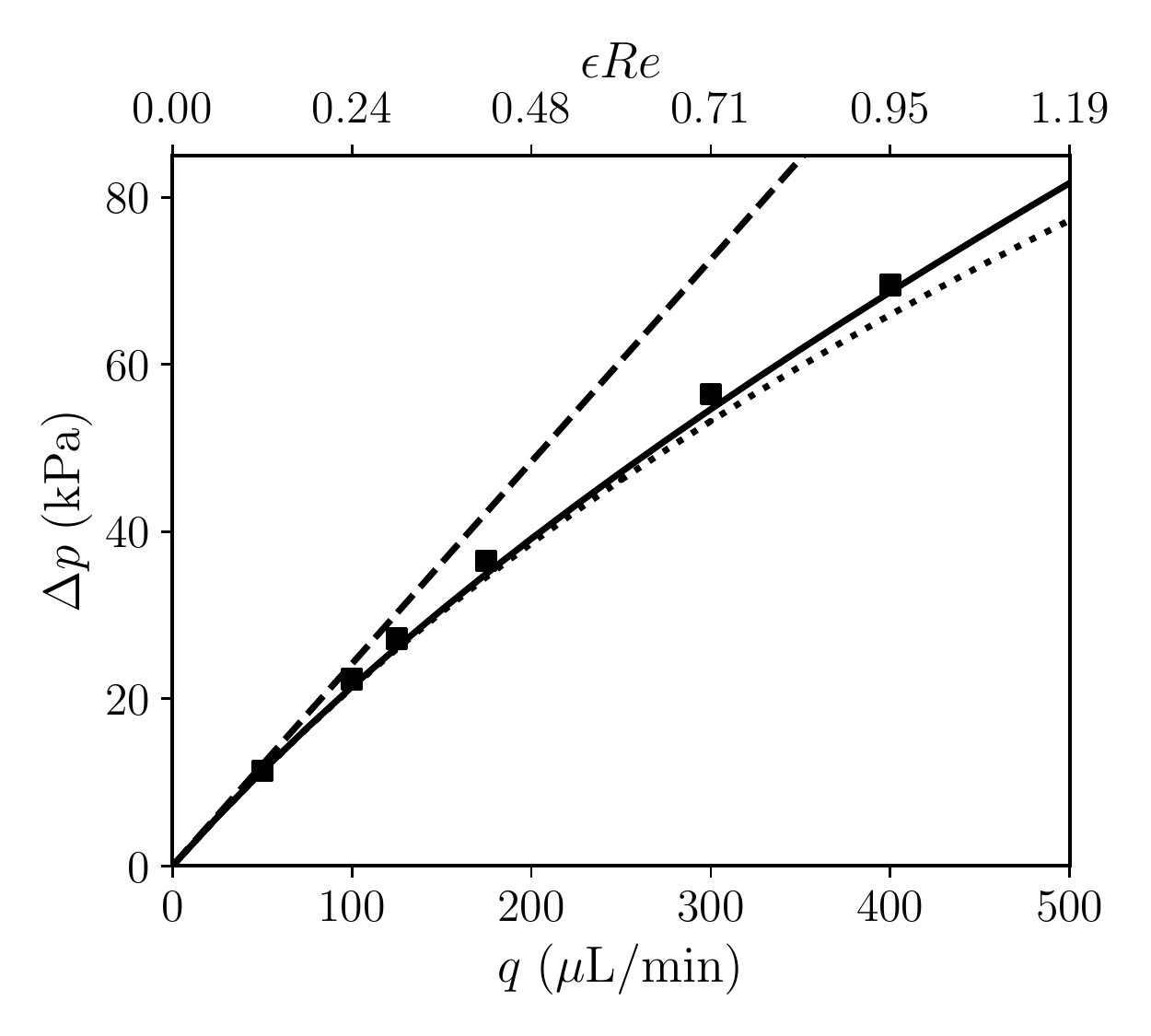}
        \label{subfig:config_b-q-dp}
    \end{subfigure}
    \begin{subfigure}[b]{0.49\textwidth}
    \centering
        \caption{}
        \includegraphics[width=.7\textwidth]{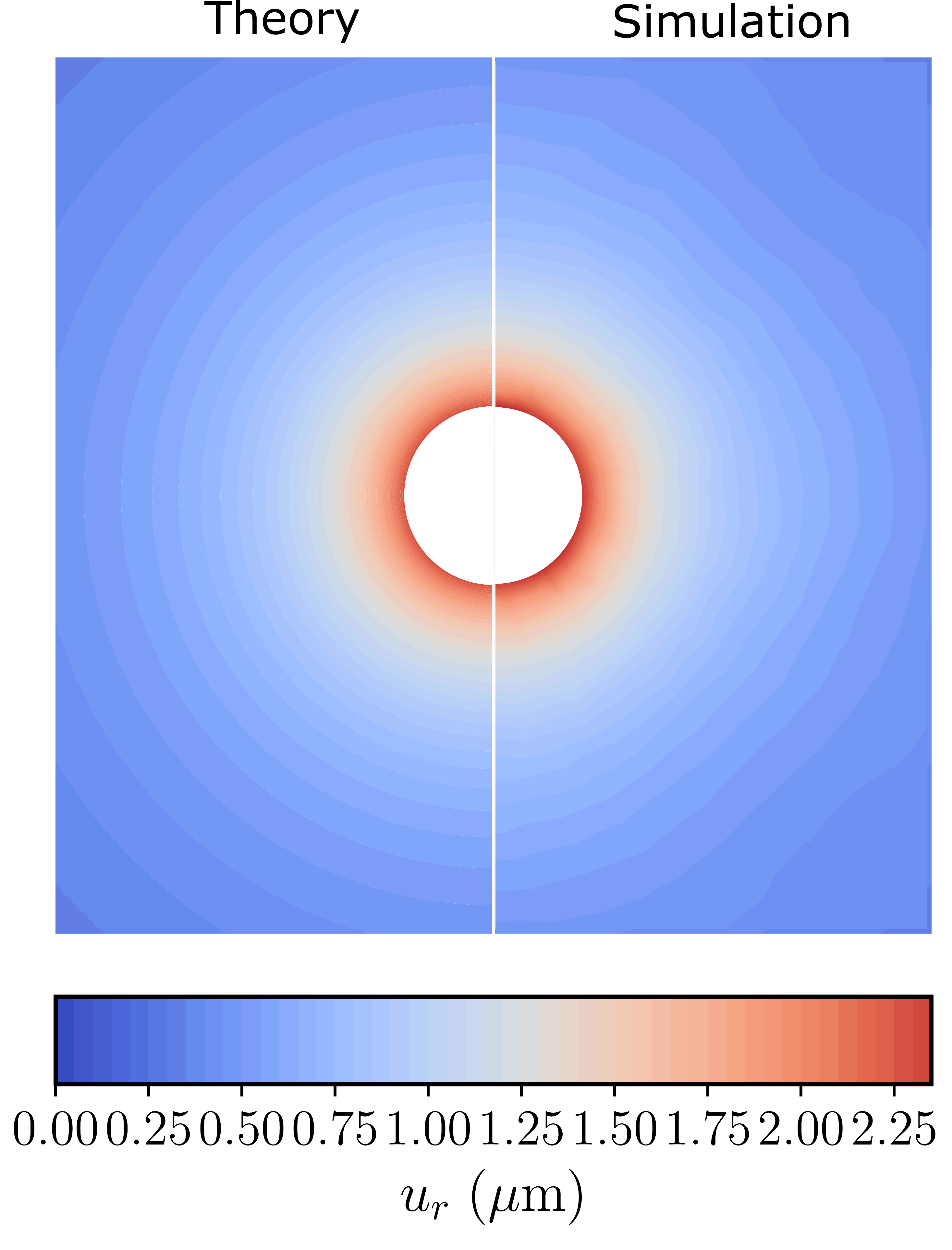}
        \label{subfig:config_b-cross}
    \end{subfigure}
    \caption{Comparison between the theory and the 3D direct simulation for configuration~II. (\textit{a}) Predicted flow rate--pressure drop relation. The solid and dotted curves represent $\Delta p$ from Eq.~\eqref{q-p-Re} and  Eq.~\eqref{q-p-2}, respectively. The dashed line represents the linear Hagen--Poiseuille law ($q$--$\Delta p$ relation) for a rigid tube (\textit{i.e.}, $\Delta p=8\mu_f q\ell/(\pi a_0^4)$). The symbols represent the simulation data. (\textit{b}) The radial displacement $u_r$ of the cross-section at $z=0.1\ell$, for $q=175~\si{\micro\litre\per\minute}$. The left half of the figure is plotted using the theory from Eq.~\eqref{q-p-Re} and Eq.~\eqref{ur-sol-b}, while the right half of the figure is plotted from the results of a SimVascular simulation.} 
    \label{fig:config_b}
\end{figure*}

\begin{figure}
    \centering
    \includegraphics[width=\linewidth]{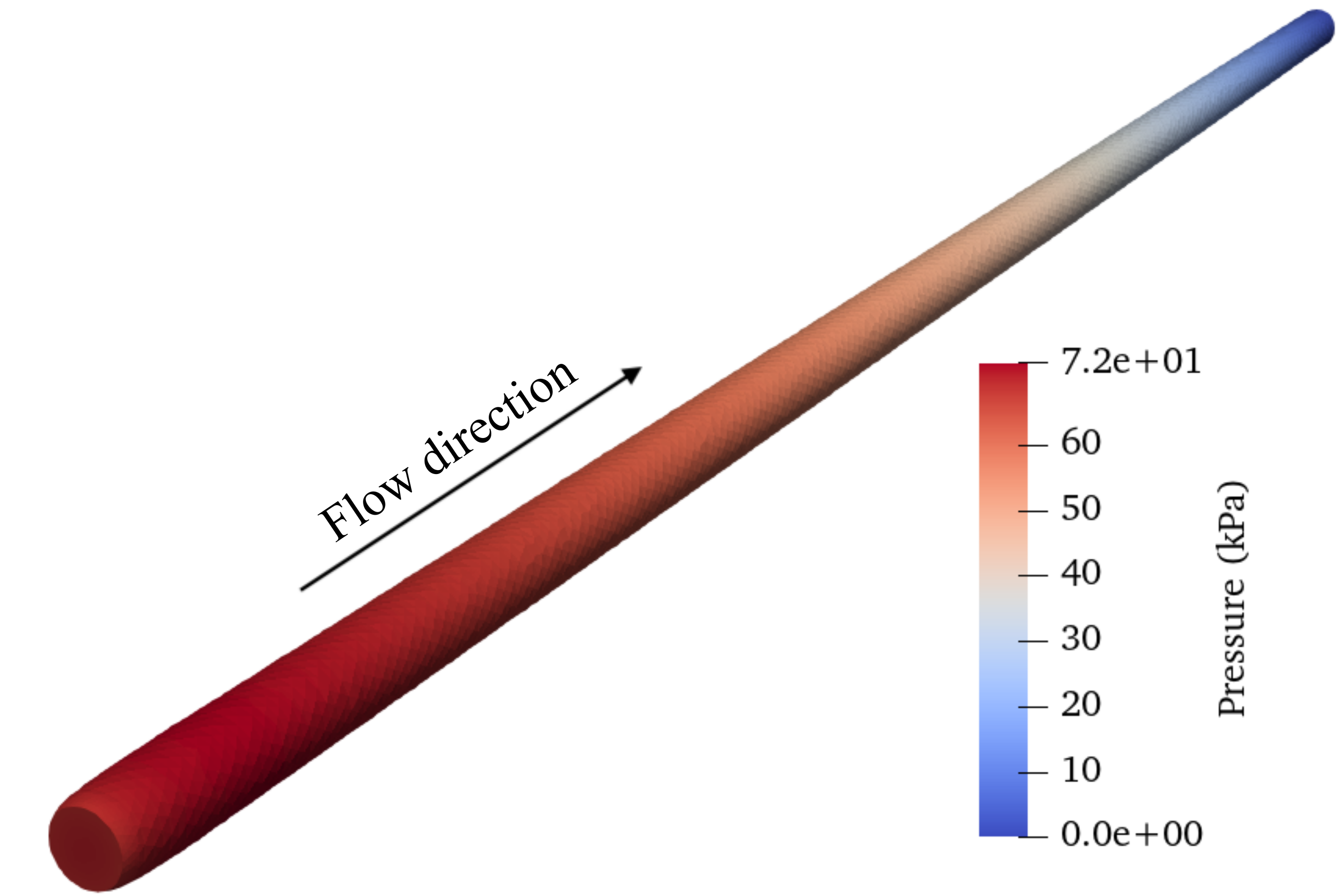}
    \caption{The deformed fluid domain of configuration II at steady state for a flow rate of $400$~\si{\micro\litre\per\minute}, obtained from a SimVascular 3D numerical simulation. The color contours show the hydrodynamic pressure distribution within the tube.}
    \label{fig:sim-screenshot}
\end{figure}

We show the results for configuration I in Fig.~\ref{fig:config_a}. From Fig.~\ref{subfig:config_a-q-dp}, we see good agreement in the value of $\Delta p$, between the theory and the 3D simulations, across a range of flow rates $q$. For larger values of $q$, $\Delta p$ predicted by Eq.~\eqref{q-p-2} for $\epsilon Re \to 0$ begins to deviate from the simulations, while $\Delta p$ predicted by Eq.~\eqref{q-p-Re}, which accounts for the inertial effects, agrees with the simulations even at $\epsilon Re = \mathcal{O}(1)$. Indeed, for $q=400$ to $500~\si{\micro\litre\per\minute}$, $\epsilon Re\approx 1$, which necessitates the inclusion of fluid inertia. 

In the inset of Fig.~\ref{subfig:config_a-q-dp}, we also show the effect of the tube wall's thickness $d$ on the flow rate--pressure drop relation. With the increase of $d$, the nonlinear variation of the pressure drop with the flow rate becomes more and more prominent. This observation can be rationalized by observing that the effective stiffness $k$ of the fluid--solid interface decreases with $d$ (recall Eq.~\eqref{a-sol-a}). Thus, a tube with larger $d$ is more prone to deformation, leading to more expansion of the cross-sectional area and, consequently, a reduction of the mean velocity (for a fixed flow rate). Therefore, the pressure losses due to viscosity are decreased, ultimately leading to a reduction in the total pressure drop (\textit{i.e.}, the total `effort' required to drive the flow) at the imposed flow rate.

{We also compare a representative predicted radial displacement of a cross-section (left half) with the corresponding 3D simulation result (right half) in Fig.~\ref{subfig:config_a-cross} at  $z=0.1\ell$. The simulated displacement at the fluid--solid interface in this cross-section is, at most, about $8.5\%$ larger than that predicted by theory.} This comparison shows satisfactory agreement, although the simulated displacement at the fluid--solid interface is slightly larger. The agreement between theory and simulation in Fig.~\ref{subfig:config_a-cross} justifies our assumptions that (i) the deformation of the tube wall is axisymmetric (ii) the 3D elasticity problem can be reduced to a 2D plane strain problem, thanks to the slenderness of the conduit's geometry. 

Overall, the results for configuration II shown in Fig.~\ref{fig:config_b} are similar to those for configuration I. Again, the flow rate--pressure drop relation predicted by Eq.~\eqref{q-p-Re} agrees better with the 3D simulations at higher flow rates, as shown in  Fig.~\ref{subfig:config_b-q-dp}, because $\epsilon Re=\mathcal{O}(1)$ for the larger flow rates imposed. {A representative predicted displacement field from Eq.~\eqref{ur-sol-b} also agrees well with the simulations, as shown in Fig.~\ref{subfig:config_b-cross}. The simulated displacement at the fluid--solid interface in the cross-section shown is, at most, about $8$\% larger than that predicted by theory.}

Importantly, although configuration II is not forced to be axisymmetric in the 3D simulations (because the outer, stress-free confinement is rectangular), the axisymmetry of the displacement field near the fluid--solid interface is maintained, consistent with the theory's assumption. Near the corners of the geometries, non-axisymmetric displacements can be discerned in the simulation plot, so there the displacement predicted by Eq.~\eqref{ur-sol-b} is no longer valid, but this has a negligible effect on the $q$--$\Delta p$ relation, as evidenced by the results in Fig.~\ref{subfig:config_b-q-dp}.

Although we have demonstrated that our theory is predictive and quantitatively accurate across a substantial range of flow rates (and, correspondingly, Reynolds numbers), some limitations should be noted. For example, small discrepancies between the theory and the 3D numerical simulation (especially in the $p(z)$ distribution) arise systematically as $q$ is increased. Of course, larger $p$ leads to larger $u_r$ in the simulations compared to the theory. This discrepancy can be partially explained by the fact that different boundary conditions are used in the theory and the simulations. In the theory, due to the plane strain reduction, we only (indirectly) imposed $u_r(a_0,\ell)=0$ (via $p(\ell)=0$) and $u_z(a_0,\ell)\equiv0$ by assumption. Meanwhile $u_r(a_0,0$) is unconstrained and $u_z(a_0,0)\equiv0$ again by assumption. Meanwhile, in the 3D simulations, \emph{both} $\bm{u}|_{z=0}=\bm{0}$ and $\bm{u}|_{z=\ell}=\bm{0}$ are imposed across the entire inlet/outlet solid regions. These conditions lead to a short inflated section of the tube near the inlet, as can be observed in Fig.~\ref{fig:sim-screenshot}. Wang and Christov \cite{WC21} rationalized this nonuniform deformation by introducing weak tension on top of the dominant plane-strain deformation. For weak finite tension, they showed that there is a positive pressure gradient in the short diverging section near the inlet when $\epsilon Re = \mathcal{O}(1)$. Therefore, {the adverse pressure gradient in the short diverging section could lead to further viscous losses, increasing the total pressure drop above the theoretical value, which may explain the remaining small discrepancies between theory and simulations in the $q$--$\Delta p$ curves in Fig.~\ref{subfig:config_a-q-dp} and Fig.~\ref{subfig:config_b-q-dp}. It would be challenging to account for the adverse pressure gradient (and the short expanding section) seen in the simulations using the theory proposed in this work, which is based on plane-strain deformation. A `boundary layer' type of calculation would have to be performed \cite{AC18b,WC21}, starting from the 3D equations of elasticity.}

In summary, our validated theory can provide important guidance for experimentalists designing new microfluidic channel configurations \cite{RCDC18,KDMCC20,KMZ21}, which can be used in reconfigurable lab-on-a-chip devices \cite{PBBK22}, for soft robotics \cite{SGOG20}, as well as related problems in the pore spaces of deformable porous media \cite{PRBM20}, such as membrane filters \cite{CLCS21}. Future work could consider shear-thinning fluids along the lines of \cite{BBG17,AC18b,PV22}, using the generalized Newtonian rheological model available in svFSI and the theory's extension to power-law and Ellis viscosity models given in \cite{C21}.

\section*{Acknowledgement}
This paper is dedicated, with admiration and respect, to Prof.\ Brian Straughan on the occasion of his 75th birthday. Specifically, I.C.C.\ would like to acknowledge Brian's keen advice throughout his career, as well as insightful and stimulating scientific discussions at numerous conferences over the years. We thank T.\ Shidhore for helpful discussions on running SimVascular simulations. We also thank E.\ Boyko for feedback on an earlier draft. Simulations were performed on the community clusters at the Rosen Center for Advanced Computing at Purdue University.

\section*{Research data availability}
SimVascular case files for the simulation, the steady-state outputs, post-processing scripts, and Jupyter notebooks implementing the theory and generating the plots in the paper are available on the Purdue University Research Repository at \url{https://purr.purdue.edu/publications/4055/}.


\balance

\appendix

\setcounter{equation}{0}
\setcounter{figure}{0}
\renewcommand{\theequation}{\thesection.\arabic{equation}}
\renewcommand{\thefigure}{\thesection.\arabic{figure}}

\section{Appendix: Flow rate--pressure drop relation at $\epsilon Re = \mathcal{O}(1)$}
\label{app:a}
For $\epsilon Re = \mathcal{O}(1)$, the flow inertia terms in Eq.~\eqref{colm-z} are no longer negligible, and the resulting equation at the leading order in $\epsilon$ cannot be integrated directly. However, $\partial p/\partial r=0$ remains true at the leading order, thus $p=p(z)$ still holds. We are thus motivated to build a 1D model by relating $p(z)$ to the flow rate $q$ \cite{PV09}. We can then make analytical progress in this case, following the derivation given in \cite{WC21} for a 2D (planar) channel.

To proceed, we assume that the axial velocity $v_z$ is still parabolic along $r$. Specifically, we invoke the von K\'arm\'an--Pohlhausen approximation \cite{W06_book,PV09,PV22}:
\begin{equation}\label{vz-q}
    v_z(r,z) = \frac{2q}{\pi a(z)^4} \left[a(z)^2-r^2\right].
\end{equation}
The above equation is obtained by making use of Eq.~\eqref{vz-O1} to eliminate $\rd p/\rd z$ from Eq.~\eqref{q-dp}.

Next, we integrate Eq.~\eqref{colm-z} over the deformed tube cross-section to obtain
\begin{multline}\label{integrate-colm-z}
    \left.2\pi\rho_f rv_r v_z\right|_{0}^{a(z)} + \int_0^{2\pi} \int_0^{a(z)} \rho_f \frac{\partial (v_z^2)}{\partial z}  r\, \rd r\, \rd \theta \\
   = \int_0^{2\pi} \int_0^{a(z)} \mu_f\frac{1}{r}\frac{\partial}{\partial r}\left(r\frac{\partial v_z}{\partial r}\right) -\frac{\rd p}{\rd z} r\,\rd r\,\rd \theta.
\end{multline}
Note that the integration of the left hand side of Eq.~\eqref{colm-z} has been simplified by integration by parts and by invoking conservation of mass, \textit{i.e.}, Eq.~\eqref{com}. Note that the first term in Eq.~\eqref{integrate-colm-z} vanishes because $\lim_{r\to0}rv_r=0$ (axisymmetry/finite flux at the centerline) and $v_z\big(a(z),z\big)=0$ (no slip). 

Next, substituting Eq.~\eqref{vz-q} into Eq.~\eqref{integrate-colm-z}, we once again obtain a first-order ODE for $p(z)$:
\begin{equation}\label{q-dp-Re}
    \frac{4\rho_f}{3\pi}\frac{\rd }{\rd z}\left[\frac{q^2}{a(z)^2}\right]=-\frac{8 q \mu_f}{a(z)^2} - \frac{\rd p}{\rd z}\pi a(z)^2.
\end{equation}
Substituting $a(z)=a_0[1+p(z)/k]$ into Eq.~\eqref{q-dp-Re} and solving the ODE subject to $p(\ell)=0$, we obtain
\begin{equation}\label{q-p-Re}
    (\ell -z )q=\frac{\pi a_0^4 k}{40\mu_f}\left\{[1+p(z)/k]^5-1\right\} - \frac{\rho_f q^2}{3\pi\mu_f}\ln[1+p(z)/k].
\end{equation}
Eq.~\eqref{q-p-Re} is an implicit equation for the pressure variation along the tube, in which the last term ($\propto q^2$ like the so-called Forchheimer correction to Darcy's law for porous media flow \cite{Bear88}) is new compared to Eq.~\eqref{q-p-1}. This new term captures the effect of finite flow inertia for $\epsilon Re = \mathcal{O}(1)$. As before, the (implicit) flow rate--pressure drop relation is obtained by taking $z=0$ in Eq.~\eqref{q-p-Re}.

\printcredits


\bibliographystyle{cas-model2-names}

\bibliography{mendeley_refs}

\begin{thebibliography}{33}
\expandafter\ifx\csname natexlab\endcsname\relax\def\natexlab#1{#1}\fi
\providecommand{\url}[1]{\texttt{#1}}
\providecommand{\href}[2]{#2}
\providecommand{\path}[1]{#1}
\providecommand{\DOIprefix}{doi:}
\providecommand{\ArXivprefix}{arXiv:}
\providecommand{\URLprefix}{URL: }
\providecommand{\Pubmedprefix}{pmid:}
\providecommand{\doi}[1]{\href{http://dx.doi.org/#1}{\path{#1}}}
\providecommand{\Pubmed}[1]{\href{pmid:#1}{\path{#1}}}
\providecommand{\bibinfo}[2]{#2}
\ifx\xfnm\relax \def\xfnm[#1]{\unskip,\space#1}\fi
\bibitem[{Anand and Christov(2021)}]{AC18b}
\bibinfo{author}{Anand, V.}, \bibinfo{author}{Christov, I.C.},
  \bibinfo{year}{2021}.
\newblock \bibinfo{title}{{Revisiting steady viscous flow of a generalized
  Newtonian fluid through a slender elastic tube using shell theory}}.
\newblock \bibinfo{journal}{Z. Angew. Math. Mech. (ZAMM)}
  \bibinfo{volume}{101}, \bibinfo{pages}{e201900309}.
\newblock \DOIprefix\doi{10.1002/zamm.201900309}.
\bibitem[{Bear(1988)}]{Bear88}
\bibinfo{author}{Bear, J.}, \bibinfo{year}{1988}.
\newblock \bibinfo{title}{{Dynamics of Fluids in Porous Media}}.
\newblock \bibinfo{publisher}{Dover Publications}, \bibinfo{address}{Mineola,
  NY}.
\bibitem[{Boyko et~al.(2017)Boyko, Bercovici and Gat}]{BBG17}
\bibinfo{author}{Boyko, E.}, \bibinfo{author}{Bercovici, M.},
  \bibinfo{author}{Gat, A.D.}, \bibinfo{year}{2017}.
\newblock \bibinfo{title}{{Viscous-elastic dynamics of power-law fluids within
  an elastic cylinder}}.
\newblock \bibinfo{journal}{Phys. Rev. Fluids} \bibinfo{volume}{2},
  \bibinfo{pages}{073301}.
\newblock \DOIprefix\doi{10.1103/PhysRevFluids.2.073301}.
\bibitem[{Boyko et~al.(2022)Boyko, Stone and Christov}]{BSC22}
\bibinfo{author}{Boyko, E.}, \bibinfo{author}{Stone, H.A.},
  \bibinfo{author}{Christov, I.C.}, \bibinfo{year}{2022}.
\newblock \bibinfo{title}{{Flow rate-pressure drop relation for deformable
  channels via fluidic and elastic reciprocal theorems}}.
\newblock \bibinfo{journal}{Phys. Rev. Fluids} \bibinfo{volume}{7},
  \bibinfo{pages}{L092201}.
\newblock \DOIprefix\doi{10.1103/PhysRevFluids.7.L092201}.
\bibitem[{Bruus(2008)}]{B08}
\bibinfo{author}{Bruus, H.}, \bibinfo{year}{2008}.
\newblock \bibinfo{title}{{Theoretical Microfluidics}}.
\newblock Oxford Master Series in Condensed Matter Physics,
  \bibinfo{publisher}{Oxford University Press}, \bibinfo{address}{Oxford, UK}.
\bibitem[{Chen et~al.(2021)Chen, Liu, Christov and Sanaei}]{CLCS21}
\bibinfo{author}{Chen, Z.}, \bibinfo{author}{Liu, S.Y.},
  \bibinfo{author}{Christov, I.C.}, \bibinfo{author}{Sanaei, P.},
  \bibinfo{year}{2021}.
\newblock \bibinfo{title}{{Flow and fouling in elastic membrane filters with
  hierarchical branching pore morphology}}.
\newblock \bibinfo{journal}{Phys. Fluids} \bibinfo{volume}{33},
  \bibinfo{pages}{062009}.
\newblock \DOIprefix\doi{10.1063/5.0054637}.
\bibitem[{Christov(2022)}]{C21}
\bibinfo{author}{Christov, I.C.}, \bibinfo{year}{2022}.
\newblock \bibinfo{title}{{Soft hydraulics: from Newtonian to complex fluid
  flows through compliant conduits}}.
\newblock \bibinfo{journal}{J. Phys.: Condens. Matter} \bibinfo{volume}{34},
  \bibinfo{pages}{063001}.
\newblock \DOIprefix\doi{10.1088/1361-648X/ac327d}.
\bibitem[{Christov et~al.(2018)Christov, Cognet, Shidhore and Stone}]{CCSS17}
\bibinfo{author}{Christov, I.C.}, \bibinfo{author}{Cognet, V.},
  \bibinfo{author}{Shidhore, T.C.}, \bibinfo{author}{Stone, H.A.},
  \bibinfo{year}{2018}.
\newblock \bibinfo{title}{{Flow rate--pressure drop relation for deformable
  shallow microfluidic channels}}.
\newblock \bibinfo{journal}{J. Fluid Mech.} \bibinfo{volume}{814},
  \bibinfo{pages}{267--286}.
\newblock \DOIprefix\doi{10.1017/jfm.2018.30}.
\bibitem[{Dillard et~al.(2018)Dillard, Mukherjee, Karnal, Batra and
  Frechette}]{DMKBF18}
\bibinfo{author}{Dillard, D.A.}, \bibinfo{author}{Mukherjee, B.},
  \bibinfo{author}{Karnal, P.}, \bibinfo{author}{Batra, R.C.},
  \bibinfo{author}{Frechette, J.}, \bibinfo{year}{2018}.
\newblock \bibinfo{title}{{A review of Winkler's foundation and its profound
  influence on adhesion and soft matter applications}}.
\newblock \bibinfo{journal}{Soft Matter} \bibinfo{volume}{14},
  \bibinfo{pages}{3669--3683}.
\newblock \DOIprefix\doi{10.1039/c7sm02062g}.
\bibitem[{Duprat and Stone(2015)}]{DS15}
\bibinfo{editor}{Duprat, C.}, \bibinfo{editor}{Stone, H.} (Eds.),
  \bibinfo{year}{2015}.
\newblock \bibinfo{title}{{Fluid-Structure Interactions in Low-Reynolds-Number
  Flows}}.
\newblock \bibinfo{publisher}{Royal Society of Chemistry},
  \bibinfo{address}{Cambridge}.
\newblock \DOIprefix\doi{10.1039/9781782628491}.
\bibitem[{Elbaz and Gat(2014)}]{EG14}
\bibinfo{author}{Elbaz, S.B.}, \bibinfo{author}{Gat, A.D.},
  \bibinfo{year}{2014}.
\newblock \bibinfo{title}{{Dynamics of viscous liquid within a closed elastic
  cylinder subject to external forces with application to soft robotics}}.
\newblock \bibinfo{journal}{J. Fluid Mech.} \bibinfo{volume}{758},
  \bibinfo{pages}{221--237}.
\newblock \DOIprefix\doi{10.1017/jfm.2014.527}.
\bibitem[{Fung(1997)}]{F97}
\bibinfo{author}{Fung, Y.C.}, \bibinfo{year}{1997}.
\newblock \bibinfo{title}{{Biomechanics: Circulation}}.
\newblock \bibinfo{edition}{2} ed., \bibinfo{publisher}{Springer-Verlag},
  \bibinfo{address}{New York, NY}.
\newblock \DOIprefix\doi{10.1007/978-1-4757-2696-1}.
\bibitem[{Gervais et~al.(2006)Gervais, El-Ali, G{\"{u}}nther and
  Jensen}]{GEGJ06}
\bibinfo{author}{Gervais, T.}, \bibinfo{author}{El-Ali, J.},
  \bibinfo{author}{G{\"{u}}nther, A.}, \bibinfo{author}{Jensen, K.F.},
  \bibinfo{year}{2006}.
\newblock \bibinfo{title}{{Flow-induced deformation of shallow microfluidic
  channels}}.
\newblock \bibinfo{journal}{Lab Chip} \bibinfo{volume}{6},
  \bibinfo{pages}{500--507}.
\newblock \DOIprefix\doi{10.1039/b513524a}.
\bibitem[{Karan et~al.(2020)Karan, Das, Mukherjee, Chakraborty and
  Chakraborty}]{KDMCC20}
\bibinfo{author}{Karan, P.}, \bibinfo{author}{Das, S.S.},
  \bibinfo{author}{Mukherjee, R.}, \bibinfo{author}{Chakraborty, J.},
  \bibinfo{author}{Chakraborty, S.}, \bibinfo{year}{2020}.
\newblock \bibinfo{title}{{Flow and deformation characteristics of a flexible
  microfluidic channel with axial gradients in wall elasticity}}.
\newblock \bibinfo{journal}{Soft Matter} \bibinfo{volume}{16},
  \bibinfo{pages}{5777--5786}.
\newblock \DOIprefix\doi{10.1039/d0sm00333f}.
\bibitem[{Kim et~al.(2021)Kim, Mitra and Zhao}]{KMZ21}
\bibinfo{author}{Kim, A.R.}, \bibinfo{author}{Mitra, S.K.},
  \bibinfo{author}{Zhao, B.}, \bibinfo{year}{2021}.
\newblock \bibinfo{title}{{Reduced Pressure Drop in Viscoelastic
  Polydimethylsiloxane Wall Channels}}.
\newblock \bibinfo{journal}{Langmuir} \bibinfo{volume}{37},
  \bibinfo{pages}{14292--14301}.
\newblock \DOIprefix\doi{10.1021/acs.langmuir.1c02087}.
\bibitem[{Lan et~al.(2018)Lan, Updegrove, Wilson, Maher, Shadden and
  Marsden}]{Lan18}
\bibinfo{author}{Lan, H.}, \bibinfo{author}{Updegrove, A.},
  \bibinfo{author}{Wilson, N.M.}, \bibinfo{author}{Maher, G.D.},
  \bibinfo{author}{Shadden, S.C.}, \bibinfo{author}{Marsden, A.L.},
  \bibinfo{year}{2018}.
\newblock \bibinfo{title}{{A Re-Engineered Software Interface and Workflow for
  the Open-Source SimVascular Cardiovascular Modeling Package}}.
\newblock \bibinfo{journal}{ASME J. Biomech. Eng.} \bibinfo{volume}{140},
  \bibinfo{pages}{024501}.
\newblock \DOIprefix\doi{10.1115/1.4038751}.
\bibitem[{Nguyen and Wereley(2006)}]{NW06}
\bibinfo{author}{Nguyen, N.T.}, \bibinfo{author}{Wereley, S.T.},
  \bibinfo{year}{2006}.
\newblock \bibinfo{title}{{Fundamentals and Applications of Microfluidics}}.
\newblock Integrated Microsystems Series. \bibinfo{edition}{2nd} ed.,
  \bibinfo{publisher}{Artech House}, \bibinfo{address}{Norwood, MA}.
\bibitem[{Paratore et~al.(2022)Paratore, Bacheva, Bercovici and
  Kaigala}]{PBBK22}
\bibinfo{author}{Paratore, F.}, \bibinfo{author}{Bacheva, V.},
  \bibinfo{author}{Bercovici, M.}, \bibinfo{author}{Kaigala, G.V.},
  \bibinfo{year}{2022}.
\newblock \bibinfo{title}{{Reconfigurable microfluidics}}.
\newblock \bibinfo{journal}{Nat. Rev. Chem.} \bibinfo{volume}{6},
  \bibinfo{pages}{70--80}.
\newblock \DOIprefix\doi{10.1038/s41570-021-00343-9}.
\bibitem[{Peir{\'{o}} and Veneziani(2009)}]{PV09}
\bibinfo{author}{Peir{\'{o}}, J.}, \bibinfo{author}{Veneziani, A.},
  \bibinfo{year}{2009}.
\newblock \bibinfo{title}{{Reduced models of the cardiovascular system}}, in:
  \bibinfo{editor}{Formaggia, L.}, \bibinfo{editor}{Quarteroni, A.},
  \bibinfo{editor}{Veneziani, A.} (Eds.), \bibinfo{booktitle}{Cardiovascular
  Mathematics}. \bibinfo{publisher}{Springer}, \bibinfo{address}{Milano}.
  volume~\bibinfo{volume}{1} of \textit{\bibinfo{series}{Modeling, Simulation
  and Applications}}. chapter~\bibinfo{chapter}{10}, pp.
  \bibinfo{pages}{347--394}.
\newblock \DOIprefix\doi{10.1007/978-88-470-1152-6_10}.
\bibitem[{Podoprosvetova and Vedeneev(2022)}]{PV22}
\bibinfo{author}{Podoprosvetova, A.}, \bibinfo{author}{Vedeneev, V.},
  \bibinfo{year}{2022}.
\newblock \bibinfo{title}{{Axisymmetric instability of elastic tubes conveying
  power-law fluids}}.
\newblock \bibinfo{journal}{J. Fluid Mech.} \bibinfo{volume}{941},
  \bibinfo{pages}{A61}.
\newblock \DOIprefix\doi{10.1017/jfm.2022.332}.
\bibitem[{Raj~M et~al.(2018)Raj~M, Chakraborty, DasGupta and
  Chakraborty}]{RCDC18}
\bibinfo{author}{Raj~M, K.}, \bibinfo{author}{Chakraborty, J.},
  \bibinfo{author}{DasGupta, S.}, \bibinfo{author}{Chakraborty, S.},
  \bibinfo{year}{2018}.
\newblock \bibinfo{title}{{Flow-induced deformation in a microchannel with a
  non-Newtonian fluid}}.
\newblock \bibinfo{journal}{Biomicrofluidics} \bibinfo{volume}{12},
  \bibinfo{pages}{034116}.
\newblock \DOIprefix\doi{10.1063/1.5036632}.
\bibitem[{Rosti et~al.(2020)Rosti, Pramanik, Brandt and Mitra}]{PRBM20}
\bibinfo{author}{Rosti, M.E.}, \bibinfo{author}{Pramanik, S.},
  \bibinfo{author}{Brandt, L.}, \bibinfo{author}{Mitra, D.},
  \bibinfo{year}{2020}.
\newblock \bibinfo{title}{{The breakdown of Darcy's law in a soft porous
  material}}.
\newblock \bibinfo{journal}{Soft Matter} \bibinfo{volume}{16},
  \bibinfo{pages}{939--944}.
\newblock \DOIprefix\doi{10.1039/C9SM01678C}.
\bibitem[{Saad(2020)}]{Saad20}
\bibinfo{author}{Saad, M.}, \bibinfo{year}{2020}.
\newblock \bibinfo{title}{{Elasticity}}.
\newblock \bibinfo{edition}{4} ed., \bibinfo{publisher}{Academic Press, an
  imprint of Elsevier}, \bibinfo{address}{London}.
\newblock \DOIprefix\doi{10.1016/C2017-0-03720-5}.
\bibitem[{Salem et~al.(2020)Salem, Gamus, Or and Gat}]{SGOG20}
\bibinfo{author}{Salem, L.}, \bibinfo{author}{Gamus, B.}, \bibinfo{author}{Or,
  Y.}, \bibinfo{author}{Gat, A.D.}, \bibinfo{year}{2020}.
\newblock \bibinfo{title}{{Leveraging viscous peeling to create and activate
  soft actuators and microfluidic devices}}.
\newblock \bibinfo{journal}{Soft Robotics} \bibinfo{volume}{7},
  \bibinfo{pages}{76--84}.
\newblock \DOIprefix\doi{10.1089/soro.2019.0005}.
\bibitem[{Shidhore and Christov(2018)}]{SC18}
\bibinfo{author}{Shidhore, T.C.}, \bibinfo{author}{Christov, I.C.},
  \bibinfo{year}{2018}.
\newblock \bibinfo{title}{{Static response of deformable microchannels: a
  comparative modelling study}}.
\newblock \bibinfo{journal}{J. Phys.: Condens. Matter} \bibinfo{volume}{30},
  \bibinfo{pages}{054002}.
\newblock \DOIprefix\doi{10.1088/1361-648X/aaa226}.
\bibitem[{Squires and Quake(2005)}]{SQ05}
\bibinfo{author}{Squires, T.M.}, \bibinfo{author}{Quake, S.R.},
  \bibinfo{year}{2005}.
\newblock \bibinfo{title}{{Microfluidics: Fluid physics at the nanoliter
  scale}}.
\newblock \bibinfo{journal}{Rev. Mod. Phys.} \bibinfo{volume}{77},
  \bibinfo{pages}{977--1026}.
\newblock \DOIprefix\doi{10.1103/RevModPhys.77.977}.
\bibitem[{Straughan(2015)}]{St15}
\bibinfo{author}{Straughan, B.}, \bibinfo{year}{2015}.
\newblock \bibinfo{title}{{Convection with Local Thermal Non-Equilibrium and
  Microfluidic Effects}}. volume~\bibinfo{volume}{32} of
  \textit{\bibinfo{series}{Advances in Mechanics and Mathematics}}.
\newblock \bibinfo{publisher}{Springer}, \bibinfo{address}{Cham, Switzerland}.
\newblock \DOIprefix\doi{10.1007/978-3-319-13530-4}.
\bibitem[{Straughan(2017)}]{St17}
\bibinfo{author}{Straughan, B.}, \bibinfo{year}{2017}.
\newblock \bibinfo{title}{{Mathematical Aspects of Multi--Porosity Continua}}.
  volume~\bibinfo{volume}{38} of \textit{\bibinfo{series}{Advances in Mechanics
  and Mathematics}}.
\newblock \bibinfo{publisher}{Springer}, \bibinfo{address}{Cham, Switzerland}.
\newblock \DOIprefix\doi{10.1007/978-3-319-70172-1}.
\bibitem[{Updegrove et~al.(2017)Updegrove, Wilson, Merkow, Lan, Marsden and
  Shadden}]{Upde17}
\bibinfo{author}{Updegrove, A.}, \bibinfo{author}{Wilson, N.M.},
  \bibinfo{author}{Merkow, J.}, \bibinfo{author}{Lan, H.},
  \bibinfo{author}{Marsden, A.L.}, \bibinfo{author}{Shadden, S.C.},
  \bibinfo{year}{2017}.
\newblock \bibinfo{title}{{SimVascular: An Open Source Pipeline for
  Cardiovascular Simulation}}.
\newblock \bibinfo{journal}{Ann. Biomed. Eng.} \bibinfo{volume}{45},
  \bibinfo{pages}{525--541}.
\newblock \DOIprefix\doi{10.1007/s10439-016-1762-8}.
\bibitem[{Wang and Christov(2019)}]{WC19}
\bibinfo{author}{Wang, X.}, \bibinfo{author}{Christov, I.C.},
  \bibinfo{year}{2019}.
\newblock \bibinfo{title}{{Theory of the flow-induced deformation of shallow
  compliant microchannels with thick walls}}.
\newblock \bibinfo{journal}{Proc. R. Soc. A} \bibinfo{volume}{475},
  \bibinfo{pages}{20190513}.
\newblock \DOIprefix\doi{10.1098/rspa.2019.0513}.
\bibitem[{Wang and Christov(2021)}]{WC21}
\bibinfo{author}{Wang, X.}, \bibinfo{author}{Christov, I.C.},
  \bibinfo{year}{2021}.
\newblock \bibinfo{title}{{Reduced models of unidirectional flows in compliant
  rectangular ducts at finite Reynolds number}}.
\newblock \bibinfo{journal}{Phys. Fluids} \bibinfo{volume}{33},
  \bibinfo{pages}{102004}.
\newblock \DOIprefix\doi{10.1063/5.0062252}.
\bibitem[{White(2006)}]{W06_book}
\bibinfo{author}{White, F.M.}, \bibinfo{year}{2006}.
\newblock \bibinfo{title}{{Viscous Fluid Flow}}.
\newblock \bibinfo{edition}{3} ed., \bibinfo{publisher}{McGraw-Hill Higher
  Education}, \bibinfo{address}{New York, NY}.
\bibitem[{Zhu et~al.(2022)Zhu, Vedula, Parker, Wilson, Shadden and
  Marsden}]{svFSI}
\bibinfo{author}{Zhu, C.}, \bibinfo{author}{Vedula, V.},
  \bibinfo{author}{Parker, D.}, \bibinfo{author}{Wilson, N.},
  \bibinfo{author}{Shadden, S.}, \bibinfo{author}{Marsden, A.},
  \bibinfo{year}{2022}.
\newblock \bibinfo{title}{{svFSI: A Multiphysics Package for Integrated Cardiac
  Modeling}}.
\newblock \bibinfo{journal}{J. Open Res. Softw.} \bibinfo{volume}{7},
  \bibinfo{pages}{4118}.
\newblock \DOIprefix\doi{10.21105/joss.04118}.

\end{thebibliography}

\end{document}